\documentclass[prl,twocolumn,showpacs,amsmath,amssymb,superscriptaddress]{revtex4}

\usepackage[dvips]{graphicx}
\usepackage{graphics}
\usepackage{dcolumn}
\usepackage{bm}
\usepackage[usenames]{color}

\begin{document}

\title{Large tunneling anisotropic magnetoresistance
in (Ga,Mn)As nanoconstrictions}

\author{A.D. Giddings}
\affiliation{School of Physics and
Astronomy, University of Nottingham, Nottingham NG7 2RD, UK}
\affiliation{Hitachi Cambridge Laboratory, Cambridge CB3 0HE, UK}

\author{M.N. Khalid}
\affiliation{Hitachi Cambridge Laboratory, Cambridge CB3 0HE, UK}

\author{J. Wunderlich}
\affiliation{Hitachi Cambridge Laboratory, Cambridge CB3 0HE, UK}

\author{S. Yasin}
\affiliation{University of Cambridge,  UK}

\author{R.P. Campion}
\affiliation{School of Physics and
Astronomy, University of Nottingham, Nottingham NG7 2RD, UK}

\author{K.W. Edmonds}
\affiliation{School of Physics and
Astronomy, University of Nottingham, Nottingham NG7 2RD, UK}

\author{J. Sinova}
\affiliation{Department of Physics, Texas A\&M University, 
College Station, TX 77843-4242, USA}

\author{T. Jungwirth}
\affiliation{Institute of Physics ASCR, Cukrovarnická 10, 162 53
Praha 6, Czech Republic} \affiliation{School of Physics and
Astronomy, University of Nottingham, Nottingham NG7 2RD, UK}
 
\author{K. Ito}
\affiliation{Hitachi Cambridge Laboratory, Cambridge CB3 0HE, UK}

\author{K. Y. Wang}
\affiliation{School of Physics and
Astronomy, University of Nottingham, Nottingham NG7 2RD, UK}

\author{D. Williams}
\affiliation{Hitachi Cambridge Laboratory, Cambridge CB3 0HE, UK}

\author{B.L. Gallagher}
\affiliation{School of Physics and
Astronomy, University of Nottingham, Nottingham NG7 2RD, UK}

\author{C.T. Foxon}
\affiliation{School of Physics and
Astronomy, University of Nottingham, Nottingham NG7 2RD, UK}
\date{\today}

\begin{abstract}
We report a large tunneling anisotropic magnetoresistance (TAMR) in a
thin (Ga,Mn)As epilayer with  lateral   nanoconstrictions. The observation establishes 
the generic nature of this effect, which originates from the spin-orbit coupling in a ferromagnet 
and  is  not specific  to  a particular  tunnel device design. The lateral geometry allows us to 
link directly normal anisotropic magnetoresistance (AMR) and TAMR. This indicates that TAMR  
may be observable in other materials showing a comparable AMR at room temperature, such as transition metal alloys.
\end{abstract}


\maketitle  

The  family of (III,Mn)V ferromagnetic semiconductors  offers unique
opportunities for  exploring the integration of two  frontier areas in
information     technologies:     spintronics    and     
nanoelectronics.  A striking example of  the synergy of the two fields
is the  very large magnetoresistance (MR) effect  recently observed in
lithographically  defined  (Ga,Mn)As  nanostructures in  which  tunnel
barriers    are   formed    in    sub-10~nm   lateral    constrictions
\cite{Ruester:2003_a}.        The      structure       studied      in
Ref.~\cite{Ruester:2003_a} consists of two such constrictions dividing
a  lithographically defined  (Ga,Mn)As wire  into contact  leads  and a
narrower central  region. The  observed $\sim 2000\%$  spin-valve like
signal was interpreted as a  type of tunneling MR (TMR) effect arising
from the relative  alignment of the magnetizations in  the regions on
either side  of the constriction, and  in which the  barrier shape was
spin dependent.  This experiment is clearly of great importance
as the size of the effect indicates that nanospintronic structures may
provide a new route to memory and sensor devices.

Recently, seemingly  unrelated strongly anisotropic hysteretic MR
of     magnitude      $\sim     3\%$     was     reported
\cite{Gould:2004_cond-mat/0407735}   in   a (Ga,Mn)As/AlOx/Au  tunneling
device. The effect is not due  to the normal TMR as only a single
ferromagnetic  layer is present.  It is a  manifestation of  a novel
tunneling  anisotropic  MR  (TAMR)  effect that  had  been  previously
overlooked.  The TAMR  arises  directly from  the  spin-orbit (SO)
coupling induced dependence of  the
tunneling  density  of  states  of  the  ferromagnetic  layer  on  the
orientation of  the magnetization with respect  to the crystallographic
axes \cite{Gould:2004_cond-mat/0407735}.  

In this  paper we report that  TAMR effects can also  dominate the MR
response of (Ga,Mn)As nanoconstrictions. It establishes that TAMR is a
generic  phenomenon 
whose occurrence is  not dependent  upon a  particular device
structure. 
The TAMR signals we observe are of order
100\%. 
We   note    that   very recent   low-temperature   studies   of
(Ga,Mn)As/GaAs/(Ga,Mn)As vertical tunnel structures find that the
TAMR can be much larger  than typical TMR signals in metallic magnetic
tunnel junctions  and astonishingly can  even lead to the  realization of a
full MR current switch \cite{Ruester:2004_cond-mat/0408532}.
Our lateral microstructures make it possible to study the link
between the normal anisotropic MR (AMR) 
\cite{Baxter:2002_a,Jungwirth:2003_b} in devices without constrictions, which also originates from
the SO coupled band structure and is present in many metallic
ferromagnets \cite{Jaoul:1977_a}, 
and TAMR measured across a tunnel junction.


\begin{figure}
\includegraphics[width=.48\textwidth]{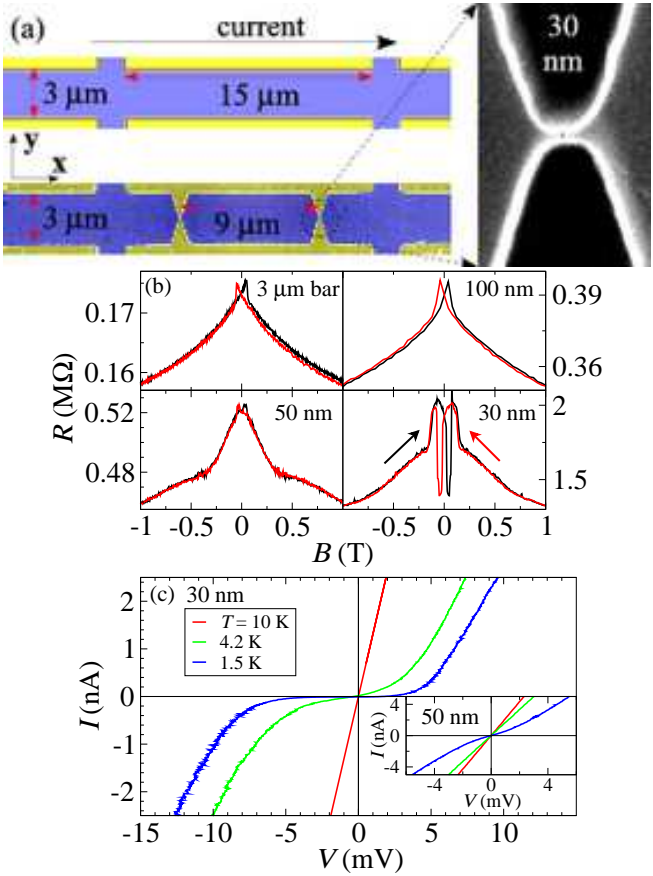}

\includegraphics[width=.41\textwidth]{Figure1bc.eps}

\caption{(a)  Schematic of  an unstructured  bar  and SEM  image of  a
double   constricted  nanodevice.     (b)  Magnetotransport
measurements  for unconstricted and  constricted devices  with applied
field  parallel  to  current at a temperature of 4.2~K (c)  I-V characteristics  for  the  30~nm
constriction device and the 50~nm device (inset)}
\label{figure1}
\end{figure}

The lateral geometry of the devices is shown in 
Fig.~\ref{figure1}(a). All microstructures discussed in this paper
were fabricated on a single Ga$_{0.98}$Mn$_{0.02}$As epilayer grown
along the [001] crystal axis
by low-temperature molecular beam epitaxy \cite{Campion:2003_a}. 
Despite
being only 5~nm thick the layer  has a Curie temperature of 40~K and
conductivity of 130~$\Omega^{-1}\,$cm$^{-1}$ at room temperature: values which are comparable
with those achieved  in high quality thicker layers for 2\%  Mn. Device fabrication
was carried out  by e-beam lithography using PMMA  positive resist and
reactive     ion    etching.    

The 3~$\mu$m wide  Hall bar,  aligned  along the [110]  direction, has  pairs of  constrictions from  30~nm to  400~nm wide separated by a  distance of 9~$\mu$m. For reference AMR experiments, a separate unstructured bar was fabricated in parallel to the stripe without constrictions.
Four point I-V curves and resistances   were  measured for both the unstructured Hall bars and across  the   constrictions (see Fig.\ref{figure1}(a)). A standard  low frequency lock-in technique was used.

The comparison of MR characteristics of different devices is presented
in Fig.~\ref{figure1}(b) for external magnetic field applied
parallel to the stripe (parallel to current). 
The unstructured bar and the 100~nm constriction show
MRs typical of the bulk (Ga,Mn)As epilayers
\cite{Baxter:2002_a,Wang:2002_a}. The  overall
isotropic (independent of applied field orientation) negative MR 
in these traces
is attributed to the suppression of magnetic disorder at large fields
\cite{Baxter:2002_a}. The hysteretic low field effect is associated
with the magnetization reversal and since its  
magnitude and sense change with applied field orientation  it is
a manifestation of the AMR. The shape of the 50~nm constriction MR partly
deviates
from this normal bulk (Ga,Mn)As behavior and a dramatic change 
is observed in the 30~nm constriction, both in the size and the sign of
the low-field effect. The marked increase of the overall
resistance of the 30~nm 
constriction device suggests that the anomalies occur due to
the formation of a tunnel junction. This is confirmed
by the measured temperature dependence of the I-V  curves. 
Constrictions
greater than  100~nm have Ohmic behavior. As shown in 
Fig.~\ref{figure1}(c), deviations  
become   more  pronounced  as  the   constriction  size  and
temperature is reduced. At low temperature and bias, conduction through
the  30~nm  constrictions is  by  tunneling.
The  occurrence of tunneling in such a wide
constriction suggests that  disorder in the very thin,  low Mn density
(Ga,Mn)As  material leads  to local  depletion  and a  tunnel barrier  of
lateral width considerably smaller than the nominal physical width.

\begin{figure}

\includegraphics[width=.41\textwidth]{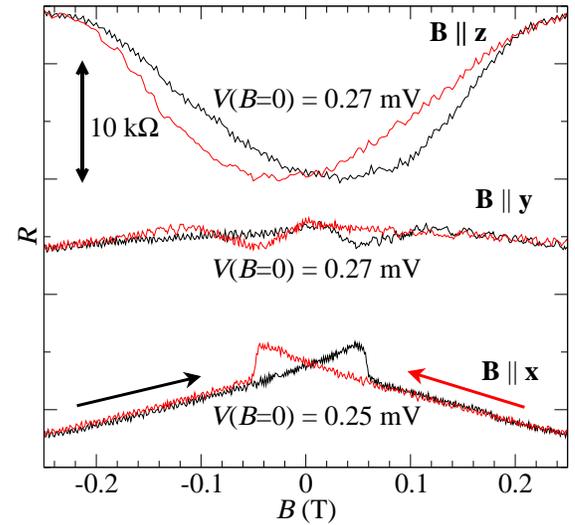}

\caption{Low field magnetotransport measurements for the unstructured
bar with applied field in three orthogonal orientations at a temperature of 4.2~K.}
\label{figure2}
\end{figure}

The negative sign of the  hysteretic effect in our tunneling device is
incompatible with TMR, for which antiparallel alignment on either side of the constriction at intermediate fields would lead to a positive hysteretic effect in the present geometry.
Instead, we interpret the data
as   the  TAMR   which  can   show  both   the  normal   and  inverted
spin-valve like  signals depending  on the  applied  field orientation
\cite{Gould:2004_cond-mat/0407735,Ruester:2004_cond-mat/0408532}.
This  interpretation  is also  consistent  with  the  geometry of  our
lateral device  in which the  central region between  constrictions and
leads  have  the  same,  relatively  large, width  and  are  therefore
expected to reverse simultaneously.

We now present a detailed analysis of the anisotropic magnetotransport
characteristics of our devices. In Fig.~\ref{figure2} we plot the low-field
AMR characteristics of the unstructured bar for magnetic fields applied parallel to the stripe (${\bf B}||{\bf x}$),
perpendicular
to the stripe in-plane (${\bf B}||{\bf y}$), 
and perpendicular to the stripe out-of-plane
(${\bf B}||{\bf z}$). The three curves in the figure are offset for clarity and
also because the absolute
comparison between resistances for different field orientations is impossible
due to our experimental set up which does not allow us to rotate the sample in 
the cryostat during the measurement. Each thermal cycling of the sample leads
to overall resistance shifts comparable to the size of the anisotropic
magnetotransport effects. Apart from 
this constant offset the MR traces are reproducible which allows us
to analyze the magnetotransport anisotropies based on the low-field parts
of individual MRs. 

We associate the hysteretic steps in the two lower curves in 
Fig.~\ref{figure2} 
with in-plane magnetization reversal precesses. A much stronger MR response is observed in the upper curve 
with the resistance increasing as the magnetization is rotated from the epilayer plane towards the vertical $z$-direction. In previously studied 50~nm thin 
Ga$_{0.98}$Mn$_{0.02}$As epilayers there was virtually no difference in the magnitude of the AMR for the two perpendicular-to-current orientations.
The large (8\%) out-of-plane AMR we observe
is therefore attributed to the
strong vertical confinement of the carriers in our ultra-thin     
Ga$_{0.98}$Mn$_{0.02}$As epilayer which breaks the symmetry between 
states with magnetization ${\bf M}||{\bf y}$ and ${\bf M}||{\bf z}$. Another indication
of confinement effects  is the presence of hysteresis in the   ${\bf B}||{\bf z}$
MR. In thicker Ga$_{0.98}$Mn$_{0.02}$As epilayers the growth direction is
magnetically hard with zero remanence due to a small compressive strain induced by the GaAs 
substrate and due to the shape anisotropy \cite{Wang:2002_a,Konig:2003_a}.
These effects compete in our epilayer with an 
increase in  the relative population  of the heavy hole  states due
to the confinement, which
tends to favor spin polarization along the growth direction \cite{Lee:2002_a}
and therefore changes the magnetic anisotropy energy landscape.

\begin{figure}

\includegraphics[width=.41\textwidth]{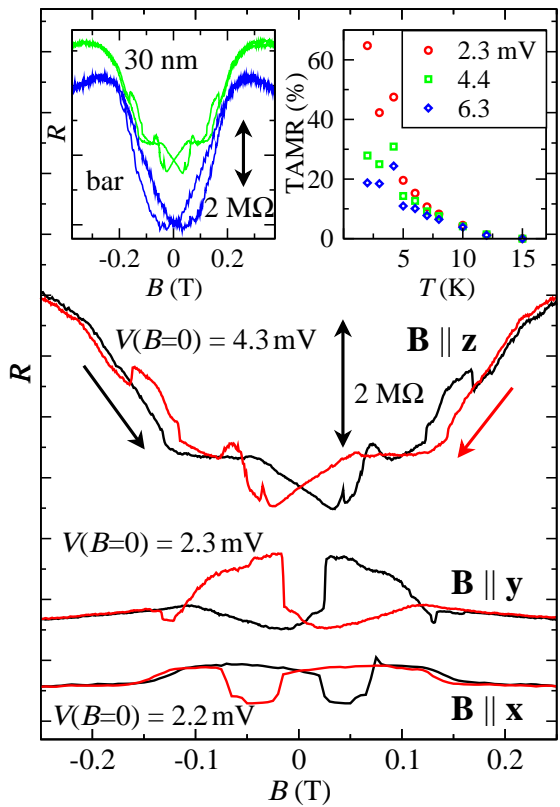}

\caption{ Detail of the TAMR measured in the 30 nm constrictions with applied field in the three orthogonal directions. Left inset: comparison of the perpendicular to plane AMR of the unconstricted stripe with the TAMR of the 30 nm constriction. The graph for the bar has been scaled up 300 times. Right inset: the temperature dependence of the TAMR  for  three  different voltages, with ${\bf B}||{\bf x}$.}
\label{figure3}
\end{figure}

The dominance of the TAMR effect in the tunneling regime is clearly demonstrated in Fig.~\ref{figure3}. This shows that the measured MR is quite different for the three orthogonal applied field directions. The comparable magnitude but opposite sign  of the TAMR for ${\bf B}||{\bf x}$ and ${\bf B}||{\bf y}$, indicates that the low resistance tunneling state is for ${\bf M}||{\bf y}$ and the high resistance state for ${\bf M}||{\bf x}$, and that the in-plane reversal process involves 90$^{\circ}$ switching through the two axes.
In the right inset of Fig.~\ref{figure3} we plot temperature dependence of the TAMR for several excitation voltages. For the lowest temperature in the figure, $T=2$~K, and lowest voltage, $V=2.3$~mV, we obtain a 65\% in-plane TAMR and the curves show no signs of saturation at these values.
Even larger TAMR signals are recorded when ${\bf M}$ is rotated out of the (Ga,Mn)As epilayer plane. For $V=4.3$~mV and $T=4.2$~K we obtained a 110\% TAMR for ${\bf M}||{\bf z}$ which compares to only 31\% for ${\bf M}||{\bf x}$ at the same temperature and excitation voltage.
 
The close correspondence between the AMR results of Fig.~\ref{figure2} and the TAMR results of Fig.~\ref{figure3} is evident. The switching events in the in-plane MR traces occur at comparable magnetic fields for the two devices. In both the AMR and the TAMR experiments, the effects at ${\bf B}||{\bf x}$ and ${\bf B}||{\bf y}$ have a similar magnitude and the opposite sign. (Note that the high and low resistance states switch places in the AMR and TAMR traces which is not surprising given the different transport regimes of the two devices.) The most important comparison
is between the ${\bf B}||{\bf z}$ AMR and TAMR as we expect the hysteretic magnetization to be unaffected by the constriction as it approaches saturation. The inset of Fig.~\ref{figure3} shows the expected similarity in general form and field scale of the AMR and TAMR in this geometry. The observation that the magnitude of the TAMR is considerably larger for ${\bf B}||{\bf z}$ than for
the in-plane fields
as is the case for the AMR, is another manifestation of the direct link between the AMR and TAMR effects. The fact that the observed TAMR effects are all much larger than the AMR effects is a manifestation of the general high sensitivity of tunneling probabilities compared to ohmic transport coefficients.

The  AMR in (Ga,Mn)As was successfully modeled \cite{Jungwirth:2003_b}
within the Boltzmann
transport theory that accounts for the SO induced anisotropies
with  respect  to the  magnetization
orientation in the hole group velocities and scattering rates.
The TAMR has  been  analyzed  in  terms of  tunneling
density  of  states anisotropies  
\cite{Gould:2004_cond-mat/0407735,Ruester:2004_cond-mat/0408532}  
or   by  calculating  the
transmission  coefficient anisotropies  using the  Landauer formalism
\cite{Brey:2004_cond-mat/0405473,Petukhov:2002_a}.    Both  approaches
confirmed  the presence  of the  TAMR effects.  The density  of states
calculations  also provided  additional qualitative  interpretation of
the  measured field-angle and  temperature dependence  of TAMR  in the
vertical                       tunnel                       structures
\cite{Gould:2004_cond-mat/0407735,Ruester:2004_cond-mat/0408532}.    
The   (Ga,Mn)As   band
structure  in   these  calculations   is  obtained  using   the  ${\bf
k}\cdot{\bf   p}$   envelope   function   description  of   the   host
semiconductor  valence   bands  in   the  presence  of   an  effective
kinetic-exchange  field produced  by  the polarized  local Mn  moments
\cite{Konig:2003_a}.

\begin{figure}[h!]
\includegraphics[angle=0,width=8cm,height=5.33cm]{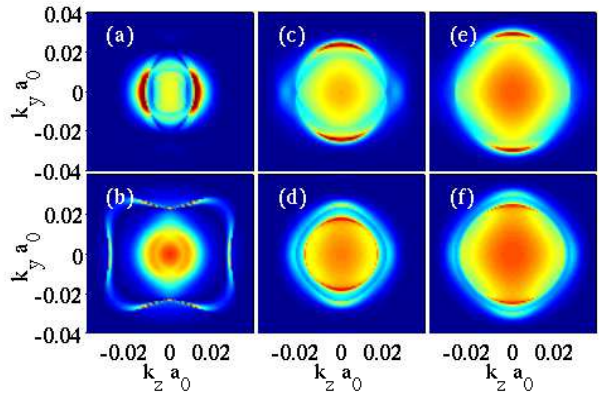}

\caption{Color   plot  of   the   calculated  tunneling   transmission
probabilities vs.  conserved in-plane  momenta at the Fermi energy.  
The  carrier densities are  $0.01$~nm$^{-3}$
(a,b), $0.05$~nm$^{-3}$ (c,d), and $0.1$~nm$^{-3}$ (e,f). 
The barrier  height and width
are 1~eV and 2~nm, respectively.  Red is the highest probability for a
given density  and blue is zero.   The tunneling current  is along the
x-direction and  the magnetization  is oriented along  the z-direction
for the  first row and  along the x-direction  for the second
row.}
\label{figure4}
\end{figure}

In  Fig.~\ref{figure4}  we   plot  illustrative  Landauer  transmission
probabilities at the Fermi energy  as a function of conserved 
momenta in the $(k_z,k_y)$-plane for 
two semi-infinite  3D (Ga,Mn)As regions separated by a
tunnel barrier. The tunnel current is along the $x$-direction.
In both ferromagnetic
semiconductor contacts we consider substitutional Mn doping
of  2\% and  a  growth direction  strain  of 0.2\%.   Details of  such
calculations can be  found in Ref.  \onlinecite{Petukhov:2002_a}.  The
additional component of  the strain, 
 which was  not considered  in previous
Landauer transport studies, allows us to model the broken cubic
symmetry      effects      observed      in     experimental      TAMR
\cite{Gould:2004_cond-mat/0407735,Ruester:2004_cond-mat/0408532}.    
The   bulk   3D   hole
densities   in     our (Ga,Mn)As   epilayer  are   of  order   1$\times
10^{20}$~cm$^{-3}$ and a gradual depletion of the carriers is expected
near the tunnel constriction. Data in panels (a) and (b) correspond to
hole density 0.1$\times 10^{20}$~cm$^{-3}$,  in (c) and (d) to density
0.5$\times  10^{20}$~cm$^{-3}$,  and  in   (e)  and  (f)  to  1$\times
10^{20}$~cm$^{-3}$.

The diagrams in  Fig.~\ref{figure4} show an intricate dependence
of the theoretical TAMR on the position in the $(k_z,k_y)$-plane.   
When integrated over
all states  at the Fermi  energy, the TAMR  ranges between
$\sim   50\%$  and  $\sim   1\%$  for   the  studied   hole  densities
0.1--1$\times  10^{20}$~cm$^{-3}$.   In  the  experimental  structure,
however, the  (Ga,Mn)As is strongly  confined in the  growth direction
which leads to depopulation of high $k_z$ momenta states. The tunnel 
constriction further reduces the number of $k_y$-states contributing to
the signal. Classically, the current is carried only by particles
with small momenta in the $x$ and $y$-directions
and wave-mechanics adds a condition $k_y=\pm\pi/w$, where $w$ is
the effective width of the constriction. 
Fig.~\ref{figure4} illustrates that 
the  theoretical TAMR can change significantly depending on the
$k_z$ and $k_y$ values selected by the confinements which suggests that 
both the  magnitude and sign of the effect are  strongly sensitive
to the detailed parameters of the tunnel barrier and of the ferromagnetic
semiconductor epilayer.

To   conclude,   we  have   established   the   TAMR   as  a   generic
effect in  tunnel devices  with  SO
coupled  ferromagnetic  contacts. The anisotropic transport nature
of the large MR signal in our lateral 
device  was demonstrated by directly comparing the TAMR 
with the AMR effects in the
contact leads.
Our measurements 
open a new avenue for integration of spintronics through the TAMR with
semiconductor nanoelectronics and motivate studies of the effect in 
other materials showing the AMR, including high Curie temperature
ferromagnetic metals.

The authors  thank L. Eaves, C. Gould, 
A.H. MacDonald,  L. Molenkamp, and P. Nov\'ak 
for useful discussions and acknowledge financial support from the
Grant  Agency of the  Czech Republic  through grant  202/02/0912,
from the EU FENIKS  project EC:G5RD-CT-2001-00535, and from
the UK EPSRC through grant GR/S81407/01.


\end{document}